\documentclass[sigconf]{acmart}

\AtBeginDocument{%
  \providecommand\BibTeX{{%
    \normalfont B\kern-0.5em{\scshape i\kern-0.25em b}\kern-0.8em\TeX}}}

\copyrightyear{2019}
\acmYear{2019}
\setcopyright{rightsretained}
\acmConference[e-Energy '19]{e-Energy '19: Proceedings of the Tenth ACM International Conference on Future Energy Systems}{June 25--28, 2019}{Phoenix, AZ, USA}
\acmBooktitle{e-Energy '19: Proceedings of the Tenth ACM International Conference on Future Energy Systems, June 25--28, 2019, Phoenix, AZ, USA}
\acmPrice{15.00}
\acmDOI{10.1145/3307772.3330158}
\acmISBN{978-1-4503-6671-7/19/06}

\usepackage{libertine}
\usepackage{txfonts}
\usepackage{lmodern}

\begin{document}

%
\title{AI Modelling and Time-series Forecasting Systems for \\ Trading Energy Flexibility in Distribution Grids}

%
\author{Bradley Eck, Francesco Fusco, Robert Gormally, Mark Purcell, Seshu Tirupathi} 
\affiliation{%
  \institution{IBM Research} 
  \streetaddress{IBM Technology Campus}   \city{Dublin}
  \state{Ireland}
}

%
\renewcommand{\shortauthors}{Eck et al.}

%
\begin{abstract}
We demonstrate progress on the deployment of two sets of technologies to support distribution grid operators 
integrating high shares of renewable energy sources, based on a market for trading local energy flexibilities.  
An artificial-intelligence (AI) grid modelling tool, based on probabilistic graphs,
predicts congestions and estimates the amount and location of 
energy flexibility required to avoid such events. A scalable time-series forecasting system 
delivers large numbers of short-term predictions of distributed energy demand and generation. 
We discuss the deployment of the technologies at three trial demonstration sites across Europe,
in the context of a research project carried out in a consortium with energy utilities, technology providers 
and research institutions.  
\end{abstract}

%
%

\ccsdesc[500]{Mathematics of computing~Time series analysis}
\ccsdesc[500]{Computing methodologies~Probabilistic reasoning}
\ccsdesc[300]{Hardware~Smart grid}

%
\keywords{}

%

%
\maketitle

\section{Introduction}


Distribution Systems Operators (DSOs) are challenged by the increasing penetration of
volatile distributed energy resources such as renewable generation and electric vehicles.
At present, in order to ensure reliable grid operation when mismatches between load and
generation patterns exist, utilities resort to curtailment or costly expansions of the existing
infrastructure. Exploitation of demand-side energy flexibility is
critical in achieving a cost-effective increase in the capacity margins of the grid. 

A set of solutions for enabling a bottom-up, market-based integration of energy
flexibility resources was developed in the context of the research project GOFLEX, funded by the European
Union and involving a consortium of energy utilities, technology providers and research
institutions across Europe \cite{goflex}. 
Residential or industrial consumers actively participate in the energy system by offering to sell 
the flexibility in their energy production and/or consumption patterns. DSOs 
offer to buy the available flexibility from the market. Traded flexibility 
is based on the concept of a flexoffer \cite{Neupane2017}. 



In this demonstration, we focus on the technology developed to automate trading decisions by 
utilities and report results from the live deployments at the three GOFLEX sites, 
in Cyprus, Switzerland and Germany. In Section \ref{sec:doms} we describe an 
artificial-intelligence (AI) grid modelling tool that determines the amount of flexibility 
({\it buy} offers) required to avoid predicted grid congestions and imbalances. Such decisions
are based on large numbers of predictions of distributed energy demand and generation, 
provided by our scalable time-series forecasting system, detailed in Section \ref{sec:forecasts}.

\section{AI-based grid modelling} \label{sec:doms}

We developed a distribution-grid observability and management system (DOMS) to 
provide decision-support to energy utilities for the prediction of localised 
congestion events and their management, by trading energy flexibility resources. 
The solution builds on a sensor data-driven model of the grid, based on 
probabilistic graphs \cite{Fusco2017,Fusco2018}. Spatio-temporal relationships 
between grid quantities are modelled with combinations of noisy sensor models and  
latent-variable neural networks, respectively defining the conditional and 
joint gaussian distributions of a factor graph. 
When energy forecasts are received (see Section \ref{sec:forecasts}), 
inference on the graph provides predictions of all the variables represented in the grid model.   
If variable predictions are outside desired operational ranges, 
inference on the graph produces estimates of the amount and location of 
energy flexibility required in order to maintain desired operation. 

\begin{figure}[h]
  \centering
  \includegraphics[width=\linewidth]{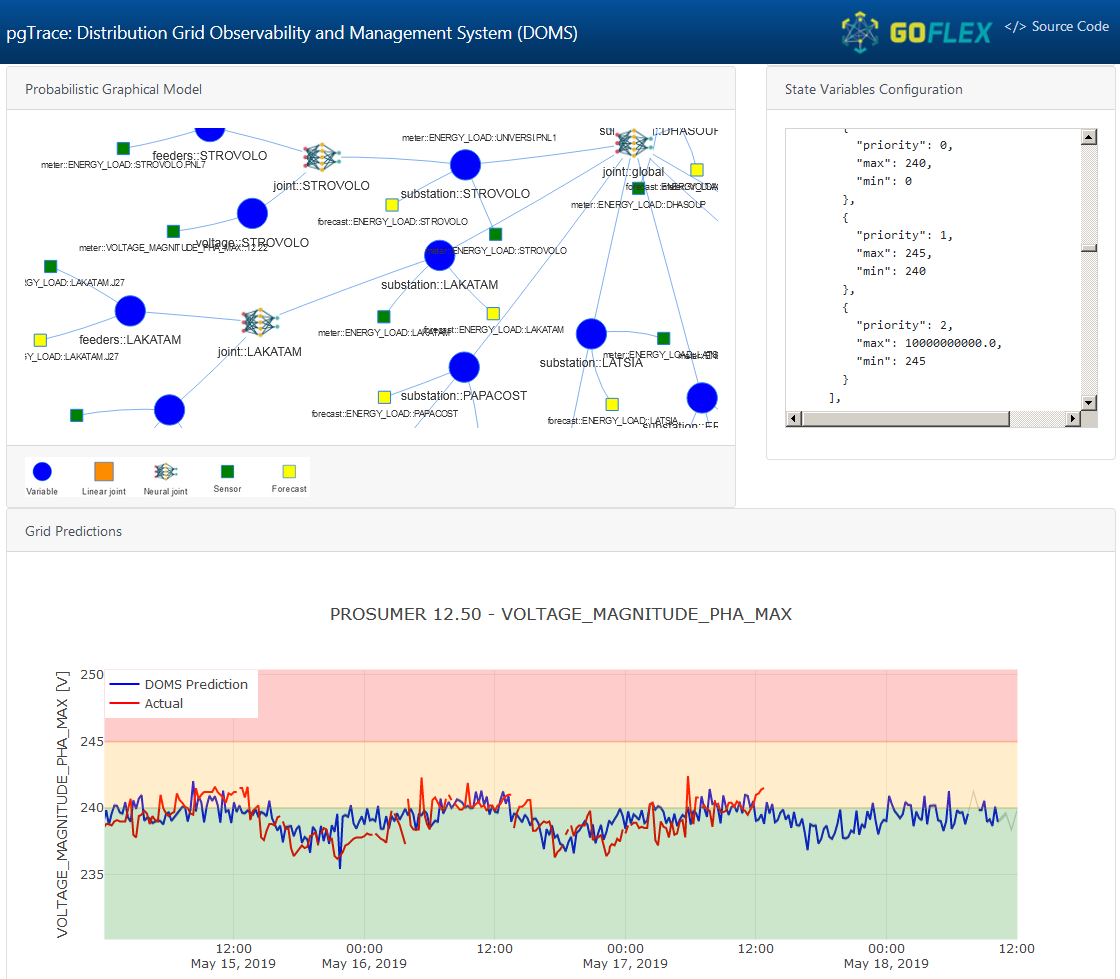}
  \caption{Grid modelling based on probabilistic graphs.}
  \label{fig:doms-screenshot}
\end{figure}

Figure \ref{fig:doms-screenshot} shows a screenshot of the DOMS graphical interface 
from the Cyprus installation. In the example, the graph models 
the load at 15 substations / 29 feeders and the voltage at 41 points, 
connected with 16 neural-network models (one for each substation, and one for the interaction between substations). 

\section{Time-series forecasting system} \label{sec:forecasts}

The grid modelling service, described in Section \ref{sec:doms}, 
relies on a large amount of sensor data and short-term forecasts 
of grid-related quantities. Depending on the specific use-case, forecasts 
might be required of energy generation from renewable sources such as wind or solar plants, 
energy load at different points in the grid, and even energy market prices. 
We developed IBM Research Castor \cite{Chen2018}, a scalable cloud-based system for 
Internet-of-Things (IoT) driven forecasts, which can easily be configured and deployed 
to serve different time-series forecasting use-cases. 
The system models the semantic context of an application, ingests data from IoT
devices and supports deployment of large numbers of custom machine-learning models 
(currently, both R or Python are supported) delivering 
live time-series forecasts according to desired training and scoring schedules.

\begin{figure}[h]
  \centering
  \includegraphics[width=\linewidth]{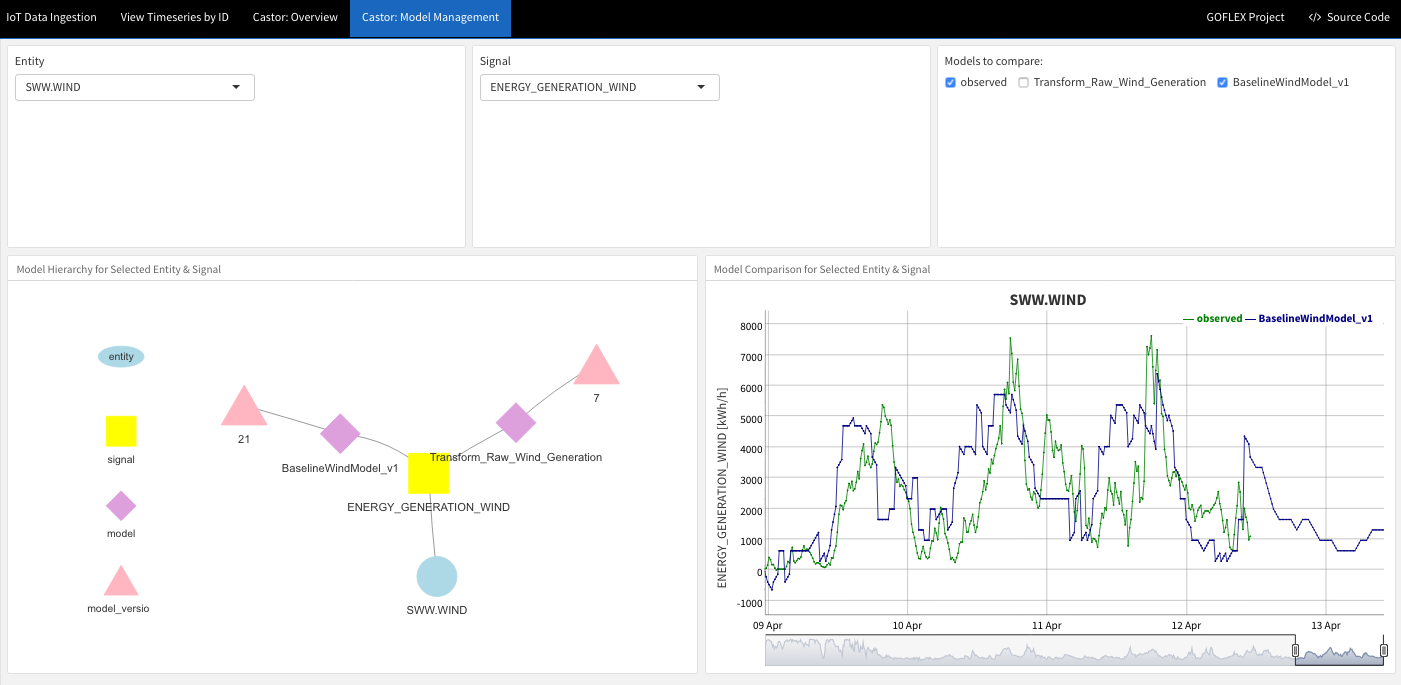}
  \caption{Time-series forecasting model management.}
  \label{fig:castor_models}
\end{figure}

A sample screenshot (Fig. \ref{fig:castor_models}) 
of the graphical interface of IBM Research Castor, from our German installation, 
demonstrates semantic-based exploration of the models and time-series forecasts deployed on the system. 
For a given semantic context, describing the application domain in terms of signals 
(what is measured, e.g. \emph{wind generation}) and entities (where, e.g. \emph{wind farm}), 
multiple models can be compared and different trained versions of the same model can be tracked. 
Similar exploration features are offered for the ingested IoT sensor data. 

Table \ref{tab:castor} summarises preliminary figures 
from the three trial installations. At the largest site, in Cyprus, 
the system handles data from 531 sensors, representing 19 different signals 
(including voltage, current, apparent power, energy) at 179 locations, 
and 174 forecasting models delivering 24-hour-ahead predictions every hour 
(at 15-minute resolution) \cite{Eck2019}. 

\begin{table}
  \caption{Forecasting system demonstrations.}
  \label{tab:sites}
  \begin{tabular}{lrrrr}
    \toprule
    & Time Series &  Entities  &  Signals  &  Models   \\
    \midrule
Germany  &      18     &       11      &       13     &        11    \\
Switzerland  &     196     &       48      &       11     &        61    \\
Cyprus  &     531     &      179      &       19     &       174    \\
\bottomrule
\end{tabular}
\label{tab:castor}
\end{table}

\section{Conclusions and Further Work} 
As trial operation on the three demonstration sites is carried out, 
the value of the proposed technologies to energy utilities will be assessed. 
In particular, the amount of flexibility that can be extracted from the energy system 
and its effectiveness in addressing grid operational issues remain open questions. 
Furthermore, the accuracy of the AI-based grid model in predicting congestion events will be evaluated. 
Finally, the scalability of the time-series forecasting system will also be further tested 
as more sensors are connected and additional prediction models are deployed. 

%
\begin{acks}
This   research   has   received   funding   from   the   European Research
Council  under  the  European  Unions  Horizon  2020 research   and innovation
programme   (grant   agreement   no.  731232). 
\end{acks}

%
\bibliographystyle{ACM-Reference-Format}
\bibliography{goflex}

%

\end{document}